# Research of Mechanical Properties of Ni-Ti-Nb Alloys on Low Temperature and Restriction Behavior

L.X. Zheng[1], N. Li[2], Y. L. Ma[1], G.M. Li[1]

(1. College of Manufacturing Science and Engineering, Southwest University of Science and Technology, Mianyang, 621010, China; 2. College of Manufacturing Science and Engineering, Sichuan University, Chengdu, 610065, China)

**Abstract:** Mechanical Properties of Ni-Ti-Nb alloys with these conditions of cold-drawing, non-vacuum heat treatment and vacuum heat treatment were measured at low temperature, and Mechanical Properties of $Ni_{47}Ti_{44}Nb_9$ alloys of restricting recover was compared with the one of alloys of non-restricting recover, and these rules of the mechanical performance between them was analyzed. Experiment indicates that, mechanical Properties of vacuum heat treatment's alloys was more excellent than the other two (non-vacuum heat treatment and cold-drawing), and the stress curves of alloys of restricting recover haven't the evident yield band, and the stress $\sigma_{0.2}$ of alloys of restricting recover was higher than the ones of alloys of non-restricting recover, but the stress $\sigma_b$ of alloys of restricting recover was lower than the ones of alloys of non-restricting recover.

The research of material mechanical properties is to study the rule of distortion and rupturing when the material is on load [1]. The guide lines to evaluate the mechanical properties of the materials include yield strength, tensile strength, elongation rate and reduction of area and so on. All these guide lines are counting bases of materials in the aspects of project apply, component design and science research etc, as well as the main basis in estimating, selecting and machining the materials [2].

The difference of recover temperature and the pretreatment temperature of the Ni-Ti-Nb alloys makes the recover state of the alloys various. Therefore, the research of alloys mechanical properties on different recover states is essential. Simultaneously, the research of the mechanical properties of Ni-Ti-Nb alloys on different pretreatment has a very important effect on its apply. This paper makes a study on the mechanical properties of the alloys at low temperature in different pretreatment, as well as the alloys of restricting recover and non-restricting recover.

## 1. Experimental material and method

The experiment material is Ni47Ti44Nb9 alloys (atomic fraction, at%), adopt the vacuum induction melting to prepare, then through hot forging, hot rolling, cold drawing to forming steel wire which has a diameter of 1mm. Take a 150mm length as a sample, non-vacuum annealing (get into the shape of a straight line in the stainless steel pipe, heats up in the resistance furnace to 900℃ and then annealing 1 hour, finally furnace cooling); or adopt vacuum annealing (get into the shape of a straight line in the quartz glass tube, and pack it after pulling out the air till 10-5 Pa vacuum, again in 900℃ annealing 1 hour, furnace cooling).

Using the JSM-5910LV scanning electron microscope and INCA Wave energy spectrometer to carries on the analysis for these test specimens of non-vacuum annealing. The mechanical properties experiment is carried on CCS-92 multi-purpose stretch testing machine reconstructed by Changchun Testing machine Research institute, and the stretch speed is 3mm/min, the transverse-load sensor shoulders 50KN.

## 2. Experimental result and analysis

### 2.1 Under different condition, the low temperature tensile curve of the alloys

Separately to carry on tensile test of test specimen in - 60℃ under the cold draw, non-vacuum annealing and vacuum annealing, and then to survey the mechanical properties parameter of these alloys under different conditions. The measurement results are presented in Figure 1

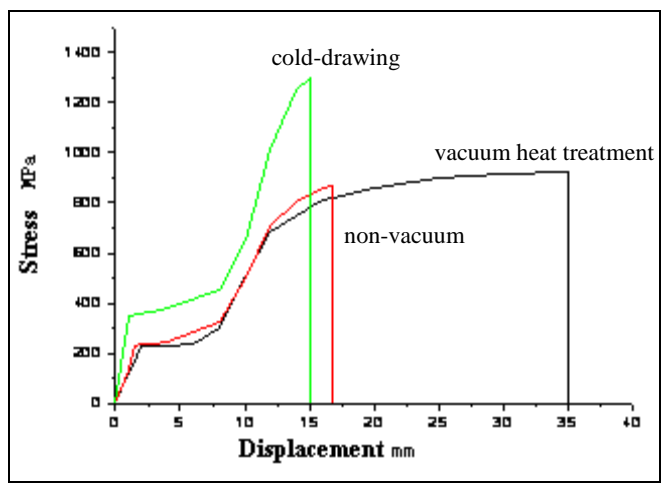

Fig.1 Stress-displacement curve of $Ni_{47}Ti_{44}Nb_9$ alloy in non-vacuum heat treatment, vacuum heat treatment and cold-drawing condition at −60℃

As indicated in Fig1, the tensile strength, yield strength and elongation rate of the test specimen under the cold drawing are $\sigma_B = 1302MPa$, $\sigma_S = 374MPa$, $\delta =$ 15% respectively. Under vacuum annealing, they are $\sigma_B = 921MPa$, $\sigma_S = 231MPa$, $\delta = 36\%$; while under non-vacuum annealing, the figures are $\sigma_B = 868MPa$, $\sigma_S = 236MPa$, $\delta = 17\%$. From the curve, we can make out that the tensile strength and the yield strength under the cold draw are obviously higher than under the two other conditions while lower in plasticity. Possibly as a result of the cold drawing way, the residual stress of parent phase in the alloy has been increased and leads to enhancing the yield strength of the stress martensitic phase. Simultaneously, the parent phase's dislocation density is increased and leads to enhancing the parent phase tensile strength which caused its elongation rate reducing. The $Ni_{47}Ti_{44}Nb_9$ alloy under non-vacuum annealing condition compared with the one under vacuum annealing condition, the both are similar in tensile strength and yield strength, but the difference between the two has doubled in plasticity. After analyzing the scanning electron microscope and energy spectrum diagram of the test specimen under non-vacuum annealing and vacuum annealing, figure 2,3, it shows that the oxygen content of the Ni47Ti44Nb9 has been increased due to the non-vacuum annealing, which results in a large number of oxygen-containing in matrix phase. While the content of Nb matrix reduced, thus the alloy's plasticity is reduced. Meanwhile, through comparison of the scanning electron microscope photographs under two different annealing of $Ni_{47}Ti_{44}Nb_9$, one can see the β-Nb particle of Ni47Ti44Nb9 alloy under vacuum annealing are obviously more than the ones under non-vacuum annealing. β-Nb particles, as a result of being soft phase, the yield stress of that is lower than the ones of the B19' martensite in Ti-Ni alloys. As uniformly distributed in the alloy and dispersion, β-Nb can enhance the plasticity of the alloy [4,5]. Therefore, the yield strength of Ni47Ti44Nb9 alloy on vacuum annealing is slightly lower than the one under non-vacuum annealing, while the elongation rate is much higher than the ones of non-vacuum annealing $Ni_{47}Ti_{44}Nb_9$ alloy.

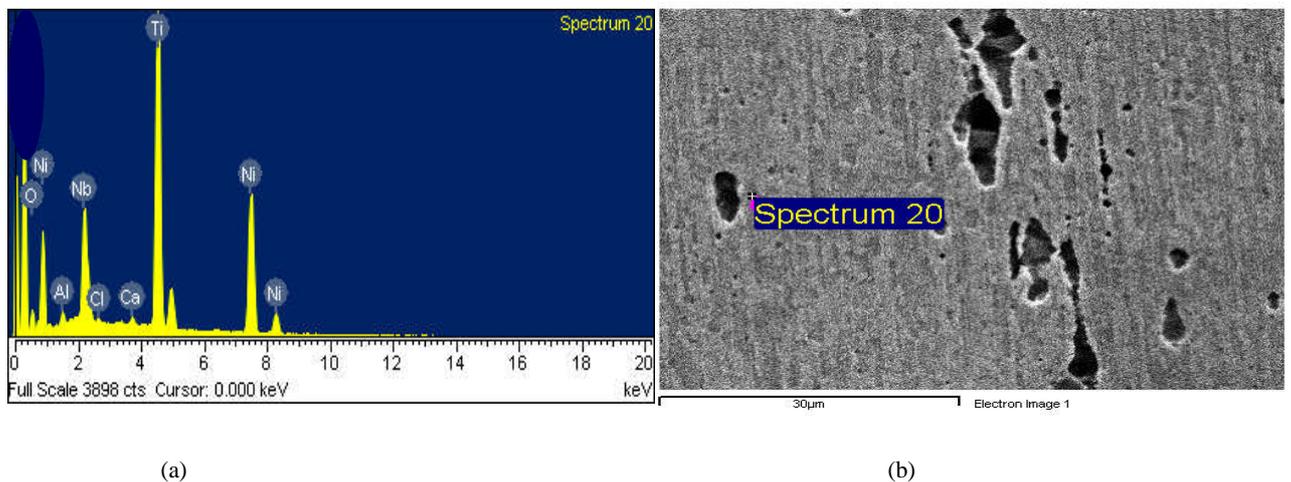

(a)            (b)

Fig.2 SEM images and Energy- Spectrum images of Ti Ni matrix in non-vacuum heat treatment condition

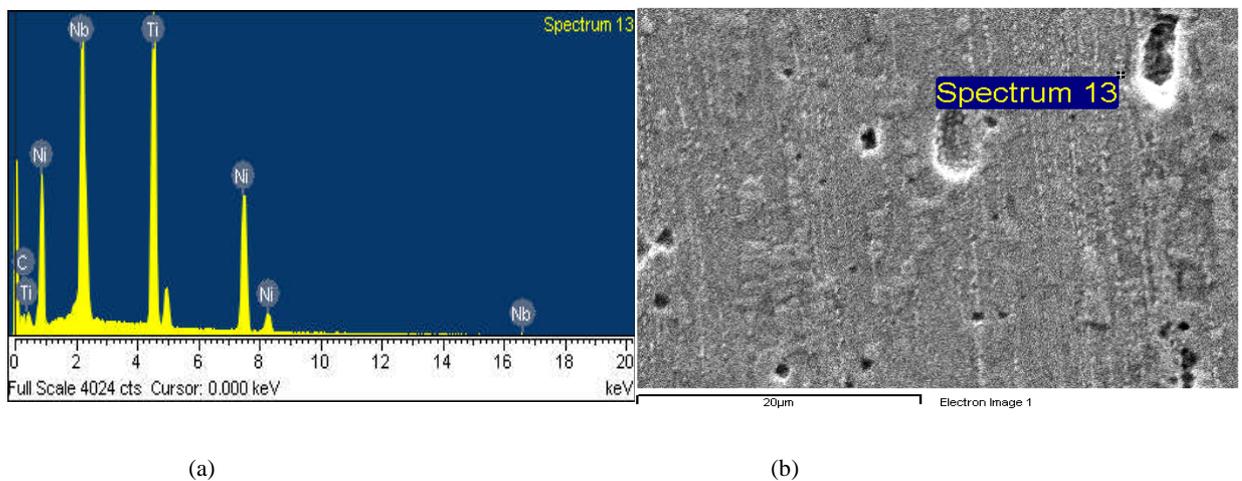

(a)            (b)

Fig.3 SEM images and Energy- Spectrum images of Ti Ni matrix in vacuum heat treatment condition

Obviously, after being managed on the condition of cold drawing and non-vacuum annealing, the elongation rate of the alloy is only about 15 ~ 17% at -60 ℃, which can not meet the requirements of engineering applications. The use of vacuum annealing has effectively prevented the oxidation and the elongation rate can reach to 36% around and that's more ideal.

## 2.2 The mechanical properties parameter of restricting recover and non-restricting recover

After 16% pre-deformation at -60℃, carry on the restricting recover for The test specimen (the test specimen clamped by stretch testing machine and heat it up to 200℃ until completely recover and cooling it to room temperature); or carry on the non-restricting recover (heat the test specimen up to 150℃ under free state until completely recover and cooling it to room temperature).

Then, carry on the mechanical properties test of the test specimen on the two conditions respectively at room temperature and form a displacement curve, as the following Fig4, (a) is Stress-displacement curve of Ni47Ti44Nb9 alloy on the condition of restricting recover and (b) is Stress-displacement curve of Ni47Ti44Nb9 alloy on the condition of non-restricting recover.

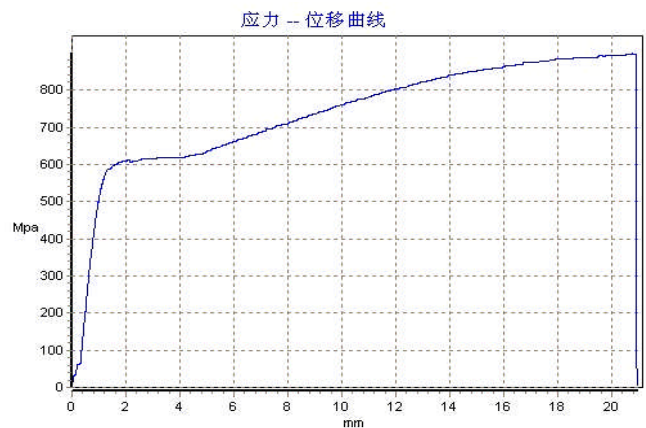

(a) At-60℃, 16% pre-deformatio, Stress-displacement curve of $Ni_{47}Ti_{44}Nb_9$ alloy on the condition of non-restricting recover

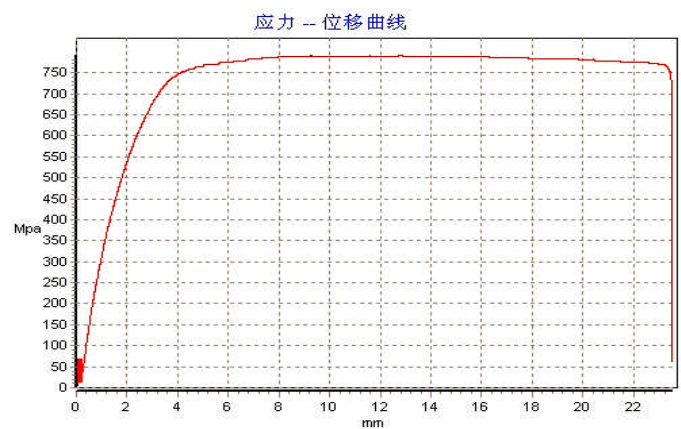

(b) At-60℃, 16% pre-deformatio , Stress-displacement curve of $Ni_{47}Ti_{44}Nb_9$ alloy on the condition of restricting recover

Fig.4 Stress-displacement curve of $Ni_{47}Ti_{44}Nb_9$ alloy on the condition of restricting recover and non-restricting recover

We can see from Fig. 4 that the tensile strength, yield strength and elongation rate of the test specimen on the condition of restricting recover are $\sigma_B = 791 MPa$ $\sigma_{0.2} = 753 MPa$, $\delta = 46\%$; while on the condition of non-restricting recover they are $\sigma_B = 899 MPa$, $\sigma_S = 620 MPa$, $\delta = 41\%$ respectively.

The figure indicate that the stress curves of alloys of restricting recover haven't the evident yield band, and the yield strength of alloys under restricting recover was higher than the ones of alloys of non-restricting recover, but the tensile strength of alloys under restricting recover was lower than the ones of alloys of non-restricting recover. This is because restricting recover causes a lot of residual dislocation in the alloy and the higher internal stress, so that the yield strength of the parent phase is increased, yet the tensile strength of the alloy actually is decreased.

## 3. Conclusion

1. The tensile strength and the yield strength on the condition of cold drawing are obviously higher than on the condition of non-vacuum annealing and vacuum annealing but lower in plasticity than the two other conditions.

2. After being managed on the condition of cold drawing and non-vacuum annealing, the elongation rate of the alloy is only about 15 ~ 17% at -60 ℃, which can not meet the requirements of engineering applications. The use of vacuum annealing has effectively prevented the oxidation and the elongation rate has reached to 36% around and that's more ideal.

3. The stress curves of alloys of restricting recover haven't the evident yield band, and the stress of alloys of restricting recover was higher than the ones of alloys of non-restricting recover, but the stress of alloys of restricting recover was lower than the ones of alloys of non-restricting recover.